# Non-uniform FIR Digital Filter Bank for Hearing Aid Application Using Frequency Response Masking Technique: A Review

*Arun Sebastian[1], Manu Francis[2], Arun Mathew[3]*

**Abstract**

Hearing aid is an electroacoustic device used to selectively amplify the audio sounds with an aim to make speech more intelligible for a hearing-impaired person. Filter bank is one of the important parts of digital hearing aid where the sub-band gains of each filter can be tuned to compensate an individual's unique hearing loss pattern. As the human perception is based on the logarithmic scale, non-uniform filter bank outperforms uniform filter bank. The main advantage of non-uniform filer bank is that it requires less number of sub-band filters, hence resulted in low hardware complexity and cost. Much effort has been devoted to design these non-uniform filter banks for hearing aid applications. This paper aimed to provide a review of previous researches based on non-uniform finite impulse response (FIR) digital filter bank for hearing aid application using frequency response masking (FRM) technique. By reviewing filter banks, we try to find the difference between fixed and variable band filter bank and to give an insight about which method is more suitable for matching most common types of hearing loss. Papers which involved methods of design, theoretical computation and simulation results of filter bank have been reviewed.

## 1. Introduction

The main task of the hearing aid is to selectively amplify the audio sounds such that the processed sound matches one's audiogram to make speech more intelligible [17, 30, 31]. The gold standard method for the diagnosis of hearing loss is pure tone audiometry which is recorded on a chart called an audiogram. The audiogram is a graph showing the audible threshold for a range of standardized frequencies. It measures the hearing threshold as a function of frequency (usually frequency as 250 / 500 / 1k / 2k / 4k / 8*kHz* in a standard audiogram). The thresholds for hearing become high at specific frequencies may cause hearing loss, i.e., they have low hearing sensitivity at specific frequencies. People with hearing loss have unique audiogram, which makes it challenging to make compensation. An ideal hearing aid should be able to adjust sound levels within a given spectrum at arbitrary frequencies. In practice, this was done by passing the input signals through a filter bank that divides the spectrum into different frequency bands, so that hearing aid can provide a precise adjustment of the gain in the required frequency [29, 35]. Therefore, hearing aid was designed to compensate for different types of hearing loss.

Based on the functionality (how the signal processed), hearing aids can be either digital or analog. Analog hearing aid functionality is much similar to an amplifier, consisting of a standard microphone to amplify sounds and then sending the signal to the speakers kept in the ear. Advancement in digital computer chips and digital signal processing method resulted in the development of digital hearing aid. Digitizing audio signal resulted in speech enhancement, less background noise, more flexibility in compensating the unique hearing loss pattern and storing multiple programs [4].

---

[1] School of Biomedical Engineering, University of Sydney, Australia, arunskumbattu@gmail.com
[2] School of Engg. & IT, UNSW Canberra, Australia, ef.manu@gmail.com
[3] Dept. of Mechanical Engg, Curtin University, Perth, Australia, arunmechme@gmail.com

Hearing losses can be compensated through tuning the sub band gains of uniform [ 27, 32] or non-uniform filter banks [1-3, 5-16, 18-22, 23-26, 33, 34, 36-51]. Usually, the mid-point value between two octave frequency points was taken as the gain of the sub band. In-order to reduce the matching error in audiogram fitting, an increased number of frequency sub bands were required because the actual hearing loss pattern (audiogram) was unique for each patient. Hence, the filter banks must have high tuning flexibility to fit various audiograms [35]. Researchers have investigated several techniques suitable for hearing aid applications. These techniques include uniform filter banks [ 27, 32], non-uniform filter banks [1-3, 7, 8, 12-16, 18-22, 24-26, 28-31, 33, 36, 38-48] and variable filter banks [5, 9-11, 28, 34 ,37, 49-51]. In uniform and non-uniform filter bank, the frequencies of the audio signal are split into different bands and then amplification was provided according to the different levels of hearing loss. In variable filter bank, the process was different and can vary gain and bandwidth (varied by changing the bandwidth ratio of sub-band filters).

The rest part of this paper will be arranged as follows: Section 2 introduces the concept and the background of finite impulse response (FIR) filters, Frequency response masking (FRM) technique, Fixed Filter Bank and Variable Filter Bank. In Section 3 reviews based on filter bank for hearing aid application using FRM Technique (different methods of designing fixed filter bank, variable filter bank and their audiogram matching results are reviewed in this section), and finally, a conclusion is given in Section 4.

## 2. Overview

2.1. FIR Filters

Filters in hearing aids can be categorized as linear or non-linear. Linear filters can amplify the audio signal by the same amount without considering the level of the signal for a given frequency while in non-linear filter amplification changes as a function amplitude of the incoming signal. FIR filters can implement linear phase filters with guaranteed stability and phase response [21] and therefore, FIR filters are highly recommended in hearing aid design. Infinite impulse response (IIR) filters can be used to design the filter bank with low computation complexity, however, the drawbacks related to non-linear phase, less stability and regular structure resulted in less acceptance in filter bank design [12]. Linear phase digital filters also have many advantages such as, free from phase distortion, and coefficient sensitivity is low under some conditions. But complexity due to the involvement of a large amount of multiplier is the main disadvantage of linear phase FIR filters [5-17, 19-23, 21, 45-48].

2.2. Non-Uniform Filter Bank

The best thing about non-uniform filter banks is that they can exhibit the same behavior as the human ear does, i.e. it can be designed logarithmically or non- uniformly. Therefore, non-uniform filter banks provide a natural means for processing audio signals. For a non-uniform filter bank, sub bands are non-uniformly spaced. In a standard audiogram hearing threshold level are performed at octave frequencies ranging from 0.5 to 8 kHz, and people with hearing impairment have unique audiogram, which makes it difficult to match the audiogram. Results suggested that the uniform filter banks may face difficulties matching the audiogram in all frequencies [ 27, 32] since sub bands in uniform filter banks are uniformly distributed throughout the entire frequency range. It worth noting that typical hearing loss such as sensorineural hearing loss (mainly caused by ageing or loud noise exposure) requires narrower bands to be assign at mid or high frequencies. Narrower bands need to be allocated to achieve better compensation at mid and high frequencies. Therefore, a non-uniform spaced digital FIR filter bank becomes very attractive [1-3, 5-16, 18-22, 23-26, 33, 34, 36-51]. In most of the cases [2, 3, 8, 12, 14, 16, 20, 21, 26, 38, 39, 41, 43, 46, 47, 48] filter bank is designed in such a way that the lower and the upper sub bands are symmetric at the mid-frequency point. These works were aimed to increase matching performance at both low and high frequencies than that of the uniform filter bank.

## 2.3 Frequency response masking (FRM) technique

The main drawback of FIR filter bank was high computational complexity caused by the design of sharp narrowband filters with very high order. This high computational complexity resulted in more power consumption and large fabrication area. In order to reduce the area and power consumption of the filter, multiplier-less architecture was introduced. Researches came up with different types of multipliers-less FIR filter architectures to enhance the speed and reduce the power consumption [2, 19, 21, 23-26, 34, 38-41].

Frequency response masking (FRM) technique is one of the most efficient way for the design of very sharp narrowband filters with a large number of sparse coefficients and widely used in hearing aid design. As this design consist of filters with large number of sparse filter coefficients, the hardware architecture required for filter implementation was very less. Also, linear phase FIR filters can be realized with guaranteed stability and phase response with FRM technique [47]. A realization structure for a filter using the FRM technique was illustrated in Fig. 1, which comprises a band-edge shaping filter $H(z)$ and two masking filters $MF(z)$ and $MF_c(z)$ to synthesize narrow transition band linear phase digital FIR filters. $H(z^M)$ was obtained by replacing each delay element of a prototype filter by $M$ delay elements and therefore had very sharp transition bands with very low arithmetic complexity. The complementary filter of $H(z^M)$ denoted by $H_c(z^M)$ can be expressed as $(z^{-\frac{N-1}{2}} - H(z^M))$ where $N$ the length of the impulse response of $H(z^M)$. Two masking filters were cascaded to $H(z^M)$ and $H_c(z^M)$, and added together to form the FRM filter with an overall transfer function [19, 24].

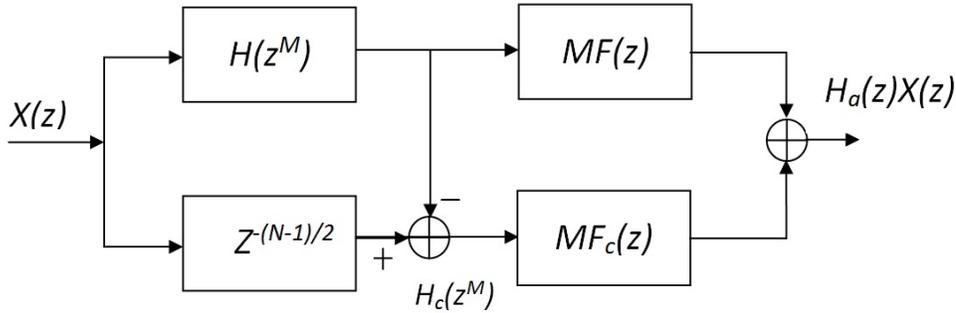

Figure 1: Realization structure of FRM filters ([Lim 1986])

The output is given by the equation

$$H_a(z) = H(z^M)MF(z) + (z^{-\frac{N-1}{2}} - H(z^M))MF_c(z)$$

$MF_c(z)$ is a complement of an original filter $MF(z)$, where $MF_c(z)$ can be implemented by subtracting the output of $MF(z)$ from the delayed version of the input. $MF_c(z)$ is given by

$$MF_c(z) = z^{-\frac{N-1}{2}} - MF(z)$$

Where N is the length of $MF(z)$. The extra delays for deriving $MF_c(z)$ from $MF(z)$ need not be implemented explicitly since the delays in $MF(z)$ can be used for this purpose. Thus, the hardware cost for producing the complementary output is minimized [23-26, 34, 38-41]. An alternative implementation of long FIR filter using FRM was proposed in [2].

A modified frequency-response masking approach was proposed in [19]. In this method, one of the sub-filters was implemented by interpolating finite impulse response (IFIR) technique. This method yields additional savings of 21% in the number of multipliers compared to the original frequency-response masking approach. An expression for the impulse response up-sampling ratio, which will minimize the complexity of the filter designed using the frequency response masking technique was implemented by Lim *et al.* [24]. Lian *et al.* introduced another modified FRM approach where one of the sub filters in FRM approach implemented by the pre-filter equalizer technique [21]. A method which uses a half-band filter to serve as one of the masking filters to achieve further saving in the number of multipliers was proposed in [25]. More reduction in the architecture complexity can be achieved by using a half-band filter for the prototype FIR filter design in FRM method. For half band filter all the odd-indexed coefficients are zero except for central coefficient and the non-zero coefficients exhibits symmetry property (as it is designed with linear phase FIR filter).

2.4. Fixed Filter Bank and Variable Filter Bank

In most of the cases, digital hearing aids adopt fixed-bandwidth filter bank for compensating hearing loss, where passband edge frequencies were fixed and the gains of each sub filters were tuned for matching the audiogram [1-3, 7, 8, 12-16, 18-22, 24-27, 39-33, 36, 38-48]. Therefore, high-accuracy audiogram fitting requires more sub filters (channels) and require more complex architecture (more additions and multiplications) which in-turn resulted in increased power consumption. However, in variable filter bank, the edge frequencies and gain of the sub filters can be tuned for audiogram fitting using sampling rate conversion technique or by transforming normalized digital filter utilizing a non-linear optimization method or Farrow method. This tuning resulted in more flexible in hearing loss compensation [5, 9-11, 28, 34,37, 49-51]. For sampling rate conversion technique, the order of the filter remains constant to achieve the change in bandwidth. The bandwidth of the fixed filter bank (designed initially) was varied by changing the bandwidth ratio using an interpolation filter or by frequency transformation of a digital filter. With this procedure a filter bank can be realized with different sub bands with varying bandwidth. However, the design process resulted in increased hardware complexity. With very low order sub filters, excellent fitting accuracy can be achieved by the variable filter bank. This type of filter banks reduces the power consumption significantly as compared with audiogram fitting using the conventional fixed-bandwidth filter bank. By using variable filter bank, we can tune the gain and edge frequency of the sub filters and its optimal values can be found to obtain the best compensation for a given audiogram/hearing loss pattern.

3. FIR Digital Filter Bank Using Frequency Response Masking Technique.

In most cases, the FRM technique is achieved by cascading different combinations of prototype filter and its interpolated versions of prototype filters to produce sub bands. The main disadvantage of linear phase FIR filter is its complexity due to the involvement of a large number of multipliers. In-order to reduce the filter complexity, filter bank uses half-band filters as prototype filters [38-41], and the sub bands are designed with symmetry at the mid-frequency point. The main advantage of a linear-phase half-band filter is the efficient implementation, which follows from two favourable properties of the filter impulse response. The first one is that the number of non-zero valued coefficients are nearly half of the filter length. The second is that non-zero coefficients exhibit symmetry property. A modified design procedure with the help of half-band filters for the frequency response masking approach was derived in [20, 25, 36]. Leading delays should be added to each filter to ensure that all sub filters have the same phase shift in order to achieve the desired frequency response and avoid frequency-dependent delay.

3.1. Fixed Filter Banks

In fixed filter banks, only the gains of the sub filters can be tuned for fitting a given audiogram therefore high accuracy audiogram fitting requires more sub filters [1-3, 7, 8, 12-16, 18-22, 24-27, 39-33, 36, 38-48]. Hearing loss can be effectively implemented by designing the filter bank structure for getting low bandwidth at regions where there is a sharp hearing loss occurs or additional bands can be introduced in the required frequency range. But this resulted in large number of sub-bands and thereby increased computation complexity and hardware cost.

A computationally efficient eight band non-uniform digital FIR filter bank for hearing aid applications was proposed in [20]. Two half-band FIR filters were employed as prototypes resulting in significant improvements in computational efficiency. The implementation of the cascaded masking filter can be effectively done through a hardware sharing scheme and developed a filter bank with the help of FRM technique and interpolation. Masking filters were used to remove the unwanted replicas from the interpolated prototype filter. A non-uniform FIR filter bank (eight sub-band filters) with stop band attenuation of $80dB$ can be implemented using 15 multipliers and ten sub filters. It gives a maximum matching error less than $4dB$ without any optimization technique (midpoint value between two successive octave frequency threshold) and less that $1.25dB$ with optimization. Optimal gains were chosen by finding the minimum least square method. The same method was adopted by Sebastian *et al.* to design an eight band non-uniform filter bank [39]. More reduction in complexity compared to the previous study [20] can be achieved by using a masking filter with minimum order since masking filter need not require very high transitional bandwidth [39]. The simulation result shows that the filter bank with stop band attenuation of $100 \ dB$ can be achieved with ten multipliers. The audiogram fitting for various hearing loss gives a maximum matching error of $2.25 \ dB$ without any optimization technique. Recently the same technique was used by Parameshap *et al.* with two equiripple FIR half-band filters for prototype filters to build a digital filter bank and achieved an average matching error of $1.38dB$ (mild hearing loss) with a stop band attenuation of $60dB$ [33].

Wei *et al.* developed a ten band non-uniform filter bank [47] which is an extension of [20] with fifteen sub filters. Two half-band filters are employed as prototypes filter, and the simulation results show that the proposed filter bank with stop band attenuation of $40dB$ can be implemented with 14 multipliers and match the hearing audiograms with less than $+/-5dB$ errors. Sebastian *et al.* studied the performance of a ten band non-uniform filter bank but with a much better result by using a masking filter with minimum order [40]. The simulation result shows that the proposed filter bank with stop band attenuation of $80dB$ can be achieved with ten multipliers. The audiogram fitting with the selected audiogram gives a maximum matching error of $2dB$. Continuing the previous work [39, 41], Sebastian *et al.* developed a 12-band filter bank and evaluated the correlation between computation complexity (number of sub band filters) and matching error [38]. The proposed method demonstrated a small reduction in error with a maximum matching error of $1.4dB$, however this required twenty-one sub filters. Furtado *et al.* developed a technique to implement a filter bank with a large number of sub band filters having sharp transition band and small roll-off factor using FRM prototype filters of cosine-modulated filter banks [8]. Comparing the results from the above studies [20, 33, 38-40, 47], it is clear that even though there is a decrease in matching error, the number of sub filters required to design the filter bank increases form ten (8 band) to twenty-one (12 band), so the complexity increases. Apart from this, the bandwidth of sub band filters is high at mid frequency and it decreases as it goes to a low and high frequency which is also symmetric with respect to mid frequency.

In most of the cases discussed above [20, 33, 38, 39, 47] narrower bands were allocated in low and high frequencies based on the fact that most common hearing problems occur in low and high frequencies. They never deal with prevalent hearing losses such as noise induced hearing loss that commonly occurs at mid frequencies (around $4kHz$). Audiogram matching for a noise induced hearing loss resulted in a matching error of $12dB$ for a filter bank with low bandwidth in low and high frequencies [40]. To accommodate these types of hearing loss, a low complex eight band non-uniform FIR digital filter bank is proposed, which provides better audiogram matching for the sharp transition of hearing loss at mid frequencies also, with ten sub bands [41]. More reduction in the number of multipliers can be achieved by designing a masking filter from the prototype filter compared to [20, 38, 39, 40, 48]. So, this design [41] requires a single half-band filter as a prototype filter, while in [20, 38, 39, 40, 48] requires two half-band filters. Also, the bandwidths of sub band filters were low in the mid frequency and increases as it goes to a lower and higher frequency. The simulation results showed that, the proposed filter bank gives 120dB attenuation with 13 multipliers and gives a maximum matching error of $3dB$ without any optimization technique. For audiogram matching the gains for each sub bands are chosen to be the midpoint values between adjacent octave thresholds in all the cases.

Devis *et al.* recommended the use of an interpolated finite impulse response (IFIR) technique to design a narrowband filter with less complexity [7]. A 17-Band non-uniform filter bank was designed with the help of a single prototype filter and by stretching the bandwidth for the desired model. This method achieved a matching error of 3.8dB for a mild hearing loss in high frequencies and 3.2dB for a mild hearing loss in all frequencies. Sharma *et al.* developed a new technique to implement an optimal filter bank using hybrid evolutionary and common sub-expression elimination technique. A prototype filter was developed with particle swarm optimization and artificial bee colony algorithms and resulted in 55% of reduction in the number of adders [42].

A 16 band non-uniform FIR digital filter bank was proposed by Wei *at al.* with the help of three prototype half-band filters [48]. Filter bank was developed with 34 multiplications and a stop band attenuation of $-60dB$, which required sixteen sub filters. This resulted in a maximum matching error of 0.4dB with normalized transition bandwidth of 0.2, which is in line with previous results [20, 33, 38-40, 47]. The major advancement compared to the previous study was the significant reduction in overall filter delay due to the novel filter structure which reduces the interpolation factor for the prototype filters. The performance was enhanced, especially in low and high frequencies, due to the increased number of filter bands.

Lee *et al.* recommended the use of Fast Filter Bank (FFB) to reduce the time delay in the filter bank [15]. Filter bank was implemented by using a single half-band FIR filter which contains very sparse coefficients and using a branching structure of cascaded sub filters. Significant reduction in delay can be achieved by using a quad-channel sub filter block in the design process. Another advantage of the FFB was that transition band of consecutive channels seamlessly added up when two consecutive channels were added to form a new channel (with a larger bandwidth). For an FFB with K number of stages, the total number of channels M, of the filter bank was: $= 2^k$. A computationally efficient FRM based FIR filter bank was introduced using two masking stages [21]. The first masking stage consists of band-edge shaping filter and its masking filter while the second comprised of complementary band-edge shaping filter and its masking filter and both of them were connected parallelly. This method demonstrated a 37% reduction in the number of multipliers compared to the conventional FRM and a 10% reduction for IFIR-FRM method. However, the group delay was slightly increased by 2%.

A 21- band ANSI S1.11 1/3 octave filter bank was proposed which consumes less power and cover the human voiced speech frequency correctly [1]. This complexity-effective multi-rate FIR filter bank algorithm reduced the power consumption by 30%-79%. The same method was adopted by Kuo *et al.* to implement an 18-band filter bank using the transformation method on multi-rate architecture to reduce the complexity [13]. This reduction complexity achieved a reduction of 16% in computation and 94% multiplications and additions compared to [1]. A 16-channel critical-like spaced, high stop band attenuation, micro power ($247.5 \, \mu W @ 1.1 \, V, 0.96 \, MHz$), small integrated circuit (IC) area ($1.62 \, mm^2 @ 0.35 - \mu m \, CMOS$) FIR filter bank core for power-critical hearing aids was proposed [6]. This high stop band attenuation was achieved by a standard pre-computational unit to generate a set of pre-calculated intermediate values that were shared by all 16 channels. Compared to a design using the usual approach, this design method resulted in reducing the power dissipation by 47% and 37% reduction in IC fabrication area. Also, the bandwidth of sub band filters increases as it goes from low frequency to high frequency. Wei et al. implemented a low-power (107.6 µW @1.1 V), small core area (2.82 mm2 @0.35 µm) filter bank (16 band) using a 0.35 µm CMOS process. Filter bank achieved a significant power saving with a stop band attenuation of $60dB$ [47].

Tang *et al.* developed a hybrid Taguchi genetic algorithm (HTGA) to optimize all coefficient values of the FRM filters simultaneously [45]. The HTGA approach is a method combining the traditional genetic algorithm, which has a robust global exploration capability, with the Taguchi method, which can exploit the optimum offspring. The high speed filter bank using a hybrid quasi-direct structure was implemented [16]. Minimal Signed Digit representation was used to express the coefficients of sub-filters, which resulted in a multiplier less architecture. In [25, 43], a low complex FRM filters using modified structure based on serial masking was proposed. The band edge shaping filter is implemented as a pre-filter equalizer cascade lowering the complexity of the band-edge shaping filter. The design of the FRM filter is formulated as a non-linear non-convex semi-infinite programming optimization problem whereby the maximum frequency response magnitude error of the overall filter is minimized. A constraint transcription method and a smoothing technique are employed to transform the continuous inequality constraints into equality constraints.

In [46], a neural network approach to FIR filter design using frequency response masking technique was proposed. The method was based on a batch back-propagation neural network algorithm (NNA), which is taken as a variable learning rate mode. In [6], non-uniform digital filter bank was proposed which uses infinite impulse response (IIR) filter as prototype filters. The design was implemented through a series of linear updates for the design variables with each update carried out using second-order cone programming. The designs of the model and masking filters were carried out utilizing semi-definite programming (SDP) and model order reduction in [6], however, in [44] the filter was optimized jointly using semi-infinite programming. In [12], the overall filter consists of a periodic model filter, its power-complementary periodic filter, and two masking filters. The model filters are composed of two all-pass filters in parallel, whereas the masking filters are linear-phase FIR filters. Computer simulations have demonstrated that the class of IIR FRM filters investigated offers an attractive alternative to its finite-impulse response compliment in terms of performance of the filter, delay of system, and realization complexity. However, guaranteed stability and phase response cannot be achieved by IIR filters.

3.2. Variable Filter Banks

In variable filter bank edge frequencies as well as their gain of the sub filters are tunable for audiogram fitting [5, 9-11, 28, 34, 37, 49-51]. As a result, this filter-bank is exceptionally flexible. In fixed filter bank different patients can't take advantage of each patients unique hearing loss. Therefore, designs which can be customized for each individual hearing-impairment become very attractive. The current trend is on implementing variable filter bank. A programmable spectrum cut-up permits to adjust filters bands relatively to patient's pathology. Programming handiness was provided to select essential speech characteristics to be considered for the patient hearing [49].

Wei *et al.* developed digital FIR filter banks with adjustable sub band distribution for hearing aids [49]. Filter bank generated 27 different sub band distribution schemes using three different prototype filters, which enables to improve the auditive performances of different individuals with unique hearing loss. The basic ideas in the design of the proposed filter bank were an interpolation, coefficient decimation and FRM technique. This method gives a matching error less than $2dB$ for a seven band filter, which is better than the eight band filter bank [20].

Yu *et al.* [51] proposed a low-complexity design of variable band edge linear phase FIR filters with sharp transition band. The filter implementation was done by decomposing the input signal into several channels in the frequency domain. The channels involved with the transition band of the variable filter due to the variation of the band edge were modeled to produce the required transition band, and then added together with all channels involved with the passband of the variable filter to make the required frequency response. The proposed variable filter has very low complexity when the transition band is sharp, if compared with other existing techniques such as the Farrow structure [11]. The computational complexity of the variable filter was even lower than that of a corresponding fixed filter with the same transition width and ripple specifications implemented in its direct form.

Bregović *et al.* developed a continuously variable bandwidth sharp FIR filter with low distortion and low complexity [5]. A fixed length FIR filter was used to build a variable filter bank with two arbitrary sampling rate converters. Arbitrary sampling rate was achieved by using a poly-phase interpolator and the fixed length filter, which was implemented by the FRM technique for reducing the complexity. The same method was used by George *et al.* in designing a filter bank, which achieved an adjacent bands rejection of 40% with continuous and enhancement of 200% in effective bandwidth better than 60dB.

George *et al.* developed a technique to design a 16-band filter bank which is non-uniformly spaced using a variable bandwidth filter to suit different varieties of hearing loss [10]. A fixed hardware was used to implement a FIR low pass filter with a fixed band width. It is then combined with two arbitrary sample rate converters to implement each sub band filters. By varying bandwidth ratio and frequency shift of each band, an optimized filter bank can be achieved for various types of hearing loss. Matching results show that it can achieve comparable results compared to the fixed filter banks for various types of hearing loss.

Haridas *et al.* adopted Farrow structure (overall response of a filter is derived as a weighted linear combination of fixed sub filters and the weights can be used to control the tunable bandwidth) method to implement a variable bandwidth filter bank [11]. Implementation was done with a set of selected bandwidth filters and frequency shifting to the selected filters with adjustable band edges makes the system reconfigurable. The audiogram fitting was tried for 4, 6, 8, and 10 bands on the sample audiograms by using the proposed method, and the results show a similar trend to previous studies and achieved a matching error between 1.5 and 2.5dB for a 10-band filter. Raghavachari *et al.* proposed a variable 6-band non-uniform digital FIR filter bank using FRM technique and approximate multiplier [34]. This method achieved an audiogram matching error of +/-4dB. Ma *et al.* developed an adjustable filter bank using Farrow method by controlling bandwidth and central frequencies and compared the results with previous studies [28]. The experiment resulted in a filter bank with more than 50% reduction in complexity and matching error reduction of 45%, 13% and 22% for 4 band, 8 band and 10 band filter banks respectively.

From the results, we can identify that for a fixed filter bank there was a decrease in matching error as the number of bands increase, but the price to pay for this was the increased number of sub filters required for the design. Therefore, the complexity increases. It is challenging to design a universal fixed filter bank that is applicable for a large number of hearing loss. For a fixed filter bank the delay introduced is significant and sometimes delays more than 20 $ms$, which resulted in hamper with lip-reading. Still, for lower matching errors, better precision in designing the filters and their cascade and parallel placements, are to be taken care of, which would increase the design cost.

While in variable filter bank the sub band filters are tuned separately to the optimum center frequencies and bandwidths to match various audiogram. Thus, once the variable filter bank is designed, the instrument can be tuned by the manufacturer to an individual user based on the audiogram characteristics. An adjustable hearing aid helps the user to adjust the device according to the change in hearing loss pattern with time or age. Yet another advantage is that the vendors of hearing aid can design an instrument to suit a set of hearing loss patterns. Here, it can be customized for any of its users, using a small set of tuning parameters. However, the adjustability of the existing filter banks is still not satisfactory. Another limitation of the variable filter bank is large hardware complexity. By the selection of the optimal number of bands, adjustability and minimum order variable band filter can reduce to the overall hardware complexity and more adjustable hearing aid.

## 4. Conclusion

Throughout this review, we considered previous research based on non-uniform FIR digital filter bank for hearing aid application using frequency response masking technique and the potential methods of designing filter bank according to literature. Hearing loss compensation can be effectively implemented by designing the filter bank structure with low bandwidth at regions where there is a sharp hearing loss occurs or additional bands can be introduced in the required frequency range. As we increase the number of bands matching error can be decreased to a certain amount, but this resulted in more number of sub bands, increased computation complexity and hardware cost. For fixed filter banks, only gain can be tuned while in variable filter bank there is a provision of changing the edge frequency and gain. Variable filter bank can be used for matching large number of hearing problems compared to the fixed filter bank as its edge frequency and gain can be adjusted. It would be more interesting if a low complex variable filter bank can be designed which can be tuned by the manufacturer to match individual audiogram characteristics with various types of hearing loss. Reflecting on the progress that has been made within this domain, it can certainly be seen that the matching performance of hearing aid achieved a significant improvement with less complex design and cost. What is particularly interesting in recent trend is that researches are aiming for a universal hearing aid with a less complicated and more adjustable variable filter bank, therefore could be applied to various type of hearing loss by tuning the parameters.

**Conflict of Interest:** *The authors have declared no conflict of interest*

## 5. References

[1] Tharini, C., Kumar, J. P., (2012, March). 21 Band 1/3-Octave Filter Bank for Digital Hearing Aids. *In Proceedings of the International Conference on Pattern Recognition, Informatics and Medical Engineering.*

[2] Bellanger, M. G. (1996, May). Improved design of long FIR filters using the frequency masking technique. In 1996 *IEEE International Conference on Acoustics, Speech, and Signal Processing Conference Proceedings* (Vol. 3, pp. 1272-1275). IEEE.

[3] Bokhari, S., & Nowrouzian, B. (2009, August). DCGA optimization of lowpass FRM IIR digital filters over CSD multiplier coefficient space. In 2009 *52nd IEEE International Midwest Symposium on Circuits and Systems* (pp. 573-576). IEEE.

[4] Bille, M., Jensen, A. M., Kjærbøl, E., Vesterager, V., Sibelle, P., & Nielsen, H. (1999). Clinical study of a digital vs an analogue hearing aid. *Scand. audiology*, 28(2), 127-135.

[5] Bregovic, R., Lim, Y. C., Saramki, T. (2008). Frequency Response Masking Based Design of Nearly Perfect Reconstruction Two Channel FIR Filterbanks with Rational Sampling Factors. *IEEE Trans. Circuits Syst. I,* 55(7), 2002-2012.

[6] Chong, K. S., Gwee, B. H., Chang, J. S. (2006). A 16-Channel Low-Power Non-uniform Spaced Filter Bank Core for Digital Hearing Aids. *IEEE Trans. Circuits Syst. II,* 53(9), 853-856.

[7] Devis, T., & Manuel, M. (2018, April). A 17-Band Non-Uniform Interpolated FIR Filter Bank for Digital Hearing Aid. In 2018 *International Conference on communication and Signal Processing* (ICCSP) (pp. 0452-0456). IEEE.

[8] Furtado, M. B., Diniz, P. S. R., Netto, S. L. (2003). Optimized Prototype Filter Based on FRM Approach for Cosine-modulated Filter Banks. *J Circ Syst Signal Pr*, 22(2), 193-210.

[9] George, J. T., Elias, E., (2012). Continuously Variable Bandwidth Sharp FIR Filters with Low Complexity. *J. Sig. Inf. Proc.*, 3, 308-315.

[10] George, J. T., Elias, E., (2014). A 16-Band Reconfigurable Hearing Aid using Variable Bandwidth Filters, *Global Journal of Research in Engineering,* 14(1).

[11] Haridas, N. T., Elias, E., (2016). Efficient variable bandwidth filters for digital hearing aid using Farrow structure, *J. Adv. Res.,* 7(2), pp. 255-262.

[12] Johansson, H., Wanhammar, L. (2000). High Speed Recursive Digital Filters Based on Frequency Response Masking Approach. *IEEE Trans. Circuits Syst. II*, 47(1), 48-61.

[13] Kuo, Y., Lin, T., Chang, W., Li, Y., Liu, C. (2008). Complexity Effective Auditory Compensation for Digital Hearing Aids. *IEEE Int. Symp. Circuits and* Systems (pp. 1472-1475). IEEE.

[14] Lee, J. W., Lim, Y. C. (2002). Efficient Implementation Of Real Filter Banks Using Frequency Response Masking Techniques. *Asia-Pacific Conf. on Circuits Syst.* (Vol. 1, pp. 69-72). IEEE.

[15] Lee, J. W., Lim, Y. C. (2008). Efficient Fast Filter Bank with a Reduced Delay. *Proc. IEEE Asia Pacific Conf. Circuits and Syst.*, 1430-1433.

[16] Lee, W. R., Caccetta, L., Teo, K. L, Rehbock, V. (2015). High Speed Multiplierless Frequency Response Masking FIR Filters with Reduced Usage of Hardware Resources. *IEEE MWSCAS* (pp. 1-4). IEEE..

[17] Levyitt, H. (1987). Digital hearing aids: Tutorial review. *J Rehabil Res Dev*, 24(4), 7-20.

[18] Lian, Y. (1995). Optimum Design of Half-band Filter Using Multi-stage Frequency Response Masking Technique. *Signal Processing J.*, 44, 369-372.
[19] Lian, Y. (2001). A New Frequency Response Masking Structure with Reduced Complexity for FIR Filter Design. *IEEE Int. Symp. Circuits Syst.*, 2, 609-612.
[20] Lian, Y., Wei, Y. (2005). Computationally Efficient Non-Uniform FIR Digital Filter Bank for Hearing Aid. *IEEE Trans. Circuits Syst.*, 52(12), 2754-2762.
[21] Lian, Y., Wei, Y. (2010). Frequency Response Masking Filters Based on Serial Masking Schemes. *J Circ Syst Signal Pr*, 29(1), 17-24.
[22] Lian, Y., Zhang, L., Ko, C. C. (2001). An Improved Frequency Response Masking Approach for Designing Sharp FIR Filters. *Signal Processing J.*, 81(12), 2573-2581.
[23] Lim, Y. C. (1986). Frequency Response Masking Approach for Synthesis of Sharp Linear Phase Digital Filters. *IEEE Trans. Circuits Syst.*, 33(4), 357-364.
[24] Lim, Y. C., Lian, Y. (1994). Frequency Response Masking Approach for Digital Filter Design: Complexity Reduction via Masking Filter Factorization. *IEEE Trans. Circuits Syst. II*, 41(8), 518-525.
[25] Lim, Y.C., Lian, Y. (1995). Reducing the Complexity of Frequency Response Masking Filters using Half-band Filters. *Signal Processing J.*, 42(2), 227-230.
[26] Liu, Y. Z., Lin, Z. P. (2008). Optimal Design of Frequency Response Masking Filters with Reduced Group Delays. *IEEE Trans. Circuits Syst. I*, 55(6), 1560-1570.
[27] Lunner, T., Hellgren, J. (1991). A Digital Filterbank Hearing Aid Design, Implementation and Evaluation. *Proc IEEE Int Conf Acoust Speech Signal Process*, 3661-3664.
[28] Ma, T., Shen, C., Wei, Y., (2019), Adjustable Filter Bank Design for Hearing Aids System, *Proc IEEE International Symposium on Circuits and Systems (ISCAS)*.
[29] Moller, A. R. (2006). Hearing: Anatomy, Physiology and Disorders of the Auditory System. Plural Publishing.
[30] Moore, B. C. J. (2007). Cochlear Hearing Loss: Physiological, Psychological and Technical Issues (2nd ed.). John Wiley & Sons.
[31] Moore, B. C. J. (2013). An Introduction to the Psychology of Hearing (6[th] ed.). *Brill*.
[32] Pantic, R. D. (2013). Design and Efficiency Analysis of one Class of Uniform Linear Phase FIR Filter Banks. *Telfor Journal*, 5(2), 165-170.
[33] Parameshap, G., Jayadevappa, D. (2018). Efficient Nonuniform Digital FIR Filter Bank Based on Masking Approach for Digital Hearing Aid. *International Conference on Intelligent Data Communication Technologies and Internet of Things (ICICI)* pp 1269-1276.
[34] Raghavachari, R., Sridharan, M., Modeling and Simulation of Frequency Response Masking FIR Filter Bank using Approximate Multiplier for Hearing Aid Application, *Adv. Syst. Sci. Appl,* 18(4), pp. 74-91.
[35] Robert, E. S. (2000). *Hearing Aid Amplification* (2[nd] ed.). Robert E. Sandlin, TX: Singular Publishers.
[36] Saramaki, T., Lim, Y. C. (1995). Synthesis of Half-Band Filters using Frequency Response Masking Technique. *IEEE Trans. Circuits Syst. II*, 42(1), 58-60.
[37] Sawant, B., Nalbalwar, S. L., Sheth, S. (2015). Design of Digital Fir Non-Uniform Reconfigurable Filter Bank for Hearing Impairments. *Int. J. Ind. Electron. Elect. Eng.,* 3, 82-85.
[38] Sebastian, A., George, J. T. (2014). A Low Complex 12-Band Non-uniform FIR Digital Filter Bank Using Frequency Response Masking Technique for Hearing Aid. *IJRITCC*, 2, 2786-2790. https://ijritcc.org/index.php/ijritcc/article/view/3296
[39] Sebastian, A., George, J. T. (2014). A Low Complex Non-uniform FIR Digital Filter Bank Using Frequency Response Masking Technique for Hearing Aid. *Proc. 15[th] National Conference on Technological Trends (NCTT),* 1384-1389.
[40] Sebastian, A., Ragesh, M. N., James, T. G. (2014, December). A low complex 10-band non-uniform FIR digital filter bank using frequency response masking technique for hearing aid. In *2014 IEEE First International Conference on Computational Systems and Communications (ICCSC)* (pp. 167-172). doi: 10.1109/COMPSC.2014.7032641.
[41] Sebastian, A., & James, T. G. (2015, February). Digital filter bank for hearing aid application using FRM technique. In *2015 IEEE International Conference on Signal Processing, Informatics, Communication and Energy Systems (SPICES)* (pp. 1-5). doi: 10.1109/SPICES.2015.7091489.
[42] Sharma, I., Kumar, A., Singh, G. K., (2017), An Efficient Method for Designing Multiplier-Less Non-Uniform Filter Bank Based on Hybrid Method Using CSE Technique, *J. Circ. Syst. Signal Pr.,* 36(3), pp. 1169–1191.

[43] Shen, T., Lim, Y. C. (2010) Recursive Digital Filter Design Based on FRM Approach. *Proc. IEEE PrimeAsia*, 217-220.

[44] Shen, T., Lim, Y. C. (2011). Low complexity Frequency Response Masking Filters Using Modified Structure Based on Serial Masking. *Proc. Eur. Signal Process. Conf.* (pp. 1400-1404). IEEE.

[45] Tang, W., Shen, T. (2010). Optimal Design of FRM Based FIR Filters by Using Hybrid Taguchi Genetic Algorithm. *Int. Conf. Green Circuits and Systems* (pp. 392-397). IEEE.

[46] Wang, X. H., He, Y. G. (2008). A Neural Network Approach to FIR Filter Design Using Frequency Response Masking Technique. *Signal Processing J.*, 88(12), 2917-2926.

[47] Wei, Y., Lian, Y. (2004). Computationally Efficient Non-Uniform Digital FIR Filter Bank for Hearing Aid. *Int. Workshop on Biomed. Circuits Syst.* (pp. S1-3). IEEE.

[48] Wei, Y., Lian, Y. (2006). A 16-Band Non-uniform FIR Digital Filter Bank for Hearing Aid. *IEEE Int. Conf. on Biomed. Circuits Syst.* (pp. 186-189). IEEE.

[49] Wei, Y., Liu, D. (2011). A Design of Digital FIR Filter Banks with Adjustable Subband Distribution for Hearing Aids. *Int. Conf. Inf Commun Sig Proc*.

[50] Wei, Y., Ma, T., Ho, B. K., Lian. Y, (2019), The Design of Low-Power 16-Band Nonuniform Filter Bank for Hearing Aids, *IEEE Trans. Biomed. Circuits Syst.,* 13(1), pp. 112-123.

[51] Yu, Y. J., Lim, Y. C., Shi, D. (2008). Low-Complexity Design of Variable Bandedge Linear Phase FIR Filters with Sharp Transition Band. *IEEE Trans. Sig. Proc.*, 57(4), 1328-1338.